\documentclass[twocolumn]{aastex631}

\newcommand{\kms}{\ensuremath{\mathrm{km\,s^{-1}}}}
\newcommand{\vsini}{\ensuremath{v\sin i}}
\newcommand{\vrot}{\ensuremath{V_\mathrm{rot}}}
\newcommand{\teff}{\ensuremath{T_\mathrm{eff}}}
\newcommand{\feh}{\ensuremath{\mathrm{[M/H]}}}
\newcommand{\logg}{\ensuremath{\log g}}
\newcommand{\msun}{\ensuremath{M_\odot}}
\newcommand{\rsun}{\ensuremath{R_\odot}}

\begin{document}

\title{The hidden companion in J1527: a 0.69 solar-mass white dwarf?}

\author[0000-0002-2419-6875]{Zhi-Xiang Zhang}
\affiliation{Department of Astronomy, Xiamen University, Xiamen, Fujian 361005, People's Republic of China}

\author[0000-0002-2912-095X]{Hao-Bin Liu}
\affiliation{Department of Astronomy, Xiamen University, Xiamen, Fujian 361005, People's Republic of China}

\author[0000-0002-5839-6744]{Tuan Yi}
\affiliation{Department of Astronomy, School of Physics, Peking University, Beijing 100871, People's Republic of China}

\author[0000-0002-0771-2153]{Mouyuan Sun}
\affiliation{Department of Astronomy, Xiamen University, Xiamen, Fujian 361005, People's Republic of China}

\author[0000-0003-3137-1851]{Wei-Min Gu}
\affiliation{Department of Astronomy, Xiamen University, Xiamen, Fujian 361005, People's Republic of China}
\correspondingauthor{Zhi-Xiang Zhang; Wei-Min Gu}
\email{zhangzx@xmu.edu.cn; guwm@xmu.edu.cn}

\begin{abstract}

Finding nearby neutron stars can probe the supernova and metal-enrichment 
histories near our Solar system. Recently, \citet{lin2023} reported an 
exciting neutron star candidate, 2MASS J15274848+3536572 (hereafter J1527), 
with a small Gaia distance of 118 parsecs. They claim that J1527 harbors an 
unseen neutron star candidate with an unusually low mass of 
$0.98\pm0.03\,M_{\odot}$. In this work, we use the 
Canada-France-Hawaii Telescope high-resolution spectrum to measure J1527's 
orbital inclination independently. Our spectral fitting suggests an orbital 
inclination of $63\pm2$ degrees. Instead, by fitting a complex hybrid 
variability model consisting of the ellipsoidal-variation component and the 
star-spot modulation to the observed light curve, \citet{lin2023} obtains 
an orbital inclination of $45.2_{-0.20}^{+0.13}$ degrees. We speculate that 
the orbital inclination obtained by the light-curve fitting is 
underestimated since J1527's light curves are obviously not pure ellipsoidal 
variations. According to our new inclination ($i\sim 63$ degrees), the mass 
of the unseen compact object is reduced to $0.69\pm0.02$\,\msun, which is as 
massive as a typical white dwarf.

\end{abstract}


\section{Introduction} \label{sec:intro}

The identification of compact objects plays a crucial role in several 
research areas, including understanding the metal enrichment process in 
the Milky Way, studying the evolution of single stars and binary systems, 
and exploring the impacts of nearby supernova explosions on the solar 
system environment. Theoretical predictions suggest that there may be 
hundreds of millions of neutron stars (NSs) in the Milky Way 
\citep[e.g.][]{Camenzind2007}. However, only a few thousand NSs have been 
found, primarily through radio pulses, and some through X-ray outbursts 
\citep{ozel2016}. The number of discoveries is far less than the 
theoretical expectations. The reason could be that the vast majority of 
NSs have weak electromagnetic radiation characteristics and are difficult 
to detect directly. To search for more compact objects in quiescent states, 
a series of works have been done based on spectroscopic or photometric 
time-domain surveys 
\citep[e.g.][]{yi2019,zheng2022,yuanhl2022,mu2022,Mazeh2022,badry2023}. 
These studies indirectly identify compact objects by analyzing the 
dynamical properties of companion stars in binary systems that contain 
unseen compact objects.

An increase of radionuclide element iron-60 ($^{60}\mathrm{Fe}$; 
half-life, 2 million years) has been detected in a sample of Pacific 
Ocean crust \citep[e.g.,][]{Koll2019,Wallner2021}. This evidence suggests 
that several nearby supernovae occurred within millions of years in the 
vicinity of the solar system. Observations have also revealed the existence 
of a local bubble (a low-density cavity of the ISM) surrounding the solar 
system \citep{Smith2001,Bergh2002,Schulreich2017}, further supporting the 
existence of nearby supernova events. Hence, it is of great interest to 
find the potential remnants, e.g., nearby NSs, of these events. Recent 
efforts in this regard include the studies of \citet{zheng2022} and 
\citet{lin2023}. \citet{zheng2022} discovered an NS candidate, J2354, 
located approximately 128 parsecs away from the solar system, in which 
the unseen compact object has a mass of $1.4\sim1.6\,\msun$. In the work 
of \citet{lin2023}, the authors claim to have identified a binary system, 
J1527, located 118 parsecs from the solar system, with an unseen neutron 
star candidate of approximately 0.98\msun. 

The mass function of the invisible star in J1527 is calculated to be 
0.131\,\msun, with the visible star being a K-type dwarf. To measure the 
mass of the unseen companion, it is vital to measure the inclination 
angle ($i$) of the binary. The inclination angle can be measured via two 
independent methods. The first method (for details, see, e.g., 
\citealt{zheng2022} and our Section \ref{sec:measure}) is to use 
high-resolution (e.g., $R > 10,000$) spectra to measure the projected 
rotation broaden velocity, $V_{\mathrm{rot}}\sin i$ (hereafter \vsini), 
of the visible star. If the visible star is tidally locked, 
$V_{\mathrm{rot}}={2\pi R_1}/{P_\mathrm{orb}}$, where $R_1$ is the visible 
star radius, and $P_\mathrm{orb}$ is the orbital period. The second method 
is to fit ellipsoidal modulations to the observed light curves. 
\citet{lin2023} adopt the latter method to estimate the inclination of 
J1527, which is about 45 degrees. However, according to TESS observations, 
the flux variations of J1527 are not dominated by ellipsoidal variations 
and other mechanisms (e.g., spots) contribute significantly to the TESS 
light curve (see figure 6 in \citealt{lin2023}). Hence, the inferred 
inclination angle in \citet{lin2023} is questionable.

In this work, we aim to independently constrain the inclination angle of 
J1574 by measuring the projected rotation broaden velocity of the visible 
star. Hence, we performed a high-resolution spectral observation by using 
the Echelle SpectroPolarimetric Device for the Observation of Stars 
(ESPaDONs) mounted on the 3.6m optical/infrared telescope at 
Canada-France-Hawaii Observatory (CFHT). We show that the inclination 
estimate based on the light-curve fitting is biased, and the actual 
inclination is $63\pm2$ degrees. As a result, the derived mass of the 
invisible compact object is significantly reduced to 
$0.69\pm0.02$\,\msun --- the typical mass of a white dwarf (WD). In 
Section \ref{sec:obs}, we introduce the observation and data reduction. 
In Section \ref{sec:measure}, we describe the measurement process of 
\vsini. The estimate of inclination and the compact object mass are 
provided in Section \ref{sec:dis}. Section \ref{sec:conc} is the 
conclusion of this paper.

\section{Observation}\label{sec:obs}

We observed J1527 twice using the ESPaDONs mounted on the CFHT telescope. 
The two observations were taken on 2023-07-09 and 2023-07-10, each 
executed with the ``object+sky'' spectroscopic mode and an exposure 
time of 660 seconds. The orbital phases of the two observations are 
0.847 and 0.756, respectively (here, we define the phase 
zero point as the visible star being at superior conjunction). The first 
observation was conducted under bad weather conditions, rendering the 
signal-to-noise ratio (SNR) of the spectrum too low to utilize. The second 
observation yielded an SNR of 20 per pixel at 6,960\,\AA, a spectral 
resolution $R\approx 69,000$, and a wavelength coverage of 
$3,700-10,500$\,\AA. Therefore, we only use the second spectrum for 
subsequent analysis.

The CFHT spectra were reduced using an automatic pipeline known as 
\texttt{Upena} (\url{https://www.cfht.hawaii.edu/Instruments/Upena/}). 
This code performs bias and flat field corrections, cosmic-ray removal, 
wavelength calibration, sky subtraction, 1D spectrum extraction, and 
heliocentric radial velocity (RV) correction. The normalized ``Star'' 
flux reduced by \texttt{Upena} is used in our following measurements. 
We use the software 
\texttt{LBLRTM}\footnote{\url{https://github.com/AER-RC/LBLRTM}}
\citep{Clough1992,Clough2005,Gullikson2014} to model and correct the 
telluric absorption. The reduced spectrum is displayed in Figure 
\ref{fig:cfht_spec}, which shows evident broaden stellar absorption lines. 
Due to the relatively low SNR of the original spectrum, it is difficult to 
clearly distinguish the absorption line profile. We resample the original 
spectrum and show it in blue lines.

\section{Measurements}\label{sec:measure}

To measure \vsini, we create a synthetic stellar spectral grid for the 
interpolation of given stellar parameters, which include the effective 
temperature (\teff), surface gravity (\logg), and metallicities (\feh). 
We employ the 
\texttt{Turbospectrum}\footnote{\url{https://github.com/bertrandplez/Turbospectrum2019}} 
\citep{Plez2012} software to generate synthetic spectra for given stellar 
parameters. For atmospheric modeling, we use the 1-D plane-parallel LTE 
MARCS model atmospheres \citep{Gustafsson2008}. The dataset of atomic and 
molecular spectral lines are from \citet{Ryabchikova2015,Heiter2021}. 
The macro-turbulence of the model is set at 2\,\kms. Because the rotational 
velocity of the visible star is significantly larger than this value, the 
macro-turbulence has a negligible impact on the fitting result. The 
wavelength interval of the synthetic spectra is set to 0.01\AA. The 
temperature grid ranges from 2700\,K to 8000\,K, the \logg\ grid spans 
from 2 to 5, and the grid for \feh\ ranges from -4 to 1. We employ the 
\texttt{scipy.interpolate.RegularGridInterpolator} function to execute 
high-dimensional interpolation on the spectral grid, thus enabling the 
acquisition of template spectra for a designated combination of \teff, 
\feh, and \logg.

In theory, stellar spectral broadening contains macro-turbulence, 
rotational broadening, and instrumental broadening. For J1527, the 
broadening of the macro-turbulence is inconsequential. The resolution 
of the CFHT spectrum corresponds to an instrumental velocity dispersion 
of ${\Delta V}_\mathrm{ins} \sim 4$\,\kms. We resample the template grid 
to match this resolution. As for the rotational broadening, we apply a 
rotational profile defined as follows:
\begin{equation}
    G(x) = \left\{ 
    \begin{array}{lr}
        \frac{2\left(1-\epsilon\right)\left(1-x^2\right)^{1/2}+\frac{\pi\epsilon}{2}\left(1-x^2\right)}{\pi\left(1-\frac{\epsilon}{3}\right)} & \mathrm{for}\ x \leq 1 \\
        0 & \mathrm{for}\ x > 1
    \end{array}
    \right.,
\end{equation}
where $x = V/\vsini$, and $\epsilon$ denotes the limb-darkening coefficient 
\citep{Gray2005}. To ascertain the limb darkening coefficient at various 
wavelengths, we refer to the tables delineated in \citet{Claret2011}. The 
observed spectrum is modeled through the convolution of the stellar 
template with the rotational broadening kernel.

We select a wavelength span from 4649\,\AA\ (35th order) to 6782\,\AA\ 
(49th order) for spectral fitting. This wavelength range has distinct 
absorption line features that can be used to constrain the \vsini. 
Additionally, there are no strong telluric absorption bands within this 
wavelength range that could impede the measurement of \vsini.

\subsection{The \vsini\ measurement with fixed temperature}

In our initial endeavor, we fix the effective temperature to 
$T_\mathrm{eff,lin} = 3917$\,K, reported by \citet{lin2023} to generate 
the template. As \citet{lin2023} does not provide additional stellar 
parameters, we measure them by minimizing the residual between the 
observation data and the template. That is, we fit \logg, \feh, and 
\vsini\ simultaneously. 

We employ the Markov Chain Monte Carlo (MCMC) to sample the posterior 
distributions of the fitting parameters. A Bayesian approach is incorporated 
to the MCMC sampler, by constructing the least square likelihood function 
and adopting priors. The likelihood function is defined as 
\begin{equation}
    \ln \mathcal{L}(\theta; O) = -\frac{1}{2}\sum_{i=1}^{n} \frac{\left[O_i - \mathrm{Model}(\theta)_i\right]^2}{\sigma_i^2},
\end{equation}
where $\mathcal{L}(\theta; O)$ is the likelihood function, $O$ is the 
observed data, and $\theta$ is the model parameters (\logg, \feh, \vsini). 
We adopt uniform priors for \feh\ and \vsini\ parameters. For surface 
gravity, we introduce a prior of $\logg = \mathcal N(4.66, 0.13)$ based 
on the stellar evolution model \citep[isochrones;][]{Morton2015}. Before 
executing the MCMC sampling, we have corrected the RV of the observed 
spectrum to the rest frame. The RV is determined based on the 
cross-correlation function (CCF) between the observed spectrum and a 
template, where the template is generated by using rough stellar 
parameters ($\teff = 4000$, $\logg = 4.7$, $\feh = -0.34$).

To conduct the MCMC fitting on the CFHT spectrum, we utilize the 
\texttt{emcee} package \citep{Foreman2013}. The MCMC program is executed 
for 10,000 iterations with 12 walkers. The autocorrelation time of the 
MCMC chain is 20. We discard the initial 89 steps when we sample the 
posterior distribution. The number of iterations is much larger than the 
autocorrection time, and the fitting results are convergent. The median, 
15.87\%, and 84.13\% quantiles of the posterior distribution are used 
as the best-fitting results, the lower, and upper $1\sigma$ uncertainties. 
The fitting result is $\vsini = 120.7_{-0.9}^{+1.0}$\,\kms, significantly 
larger than the estimate of ${\vsini}_\mathrm{lin} = 94\pm 5$\,\kms\ reported 
in \citet{lin2023}. The fitting result with a fixed temperature is listed 
in Table \ref{tab:CFHT_info}. The chi-square value between the spectrum 
and the best-fitting template is $\chi^2 = 103,203.6$, corresponding to a 
reduced-$\chi^2$ of 1.294 (with the degree of freedom 
$N_{\mathrm{dof}}=79,753$).

In a subsequent experiment, we fix the stellar parameters to their 
best-fit values, but change \vsini\ to $94\,\kms$ as reported by 
\citet{lin2023}, and generate a new template spectrum. The chi-square 
between this regenerated template and the observed data is 
$\chi^2_\mathrm{lin} = 103,994.7$, corresponding to a 
reduced-$\chi^2_\mathrm{lin}$ of 1.304. The 
$\chi^2_\mathrm{lin} = 103,994.7$ is larger than that of our best fit, 
$\chi^2 = 103,203.6$ by a value of nearly 700, clearly demonstrating 
the superiority of our fitting result.

\subsection{The \vsini\ measurement with free temperature}

To circumvent potential \vsini\ measurement bias arising from the mismatch 
between the template and observed data, we now also free temperature, to 
conduct a full parameter fitting. In this configuration, the fitting result 
is $\vsini = 122.5_{-1.0}^{+1.1}$\,\kms. This newly determined \vsini\ is 
consistent with our previous result (see also Table \ref{tab:CFHT_info}; 
right column). The chi-square of this fitting procedure is 
$\chi^2 = 102,916.4$, corresponding to a reduced-$\chi^2$ of 1.290.

The aforesaid two fitting outcomes are proximate. To ascertain which 
constitutes our definitive conclusion, we turn to the Akaike information 
criterion (AIC) and Bayesian information criterion (BIC) for comparison 
\citep{Stoica2004}. For the fitting process with a fixed temperature, 
the AIC and BIC of the fitting are 103,211 and 103,239, respectively. For 
the fitting model with a free effective temperature, the AIC and BIC are 
102931 and 102968, respectively. Hence, we adopt 
$\vsini = 122.5_{-1.0}^{+1.1}$\,\kms\ as our final result.

\begin{figure*}
    \centering
    \includegraphics[width=0.9\textwidth]{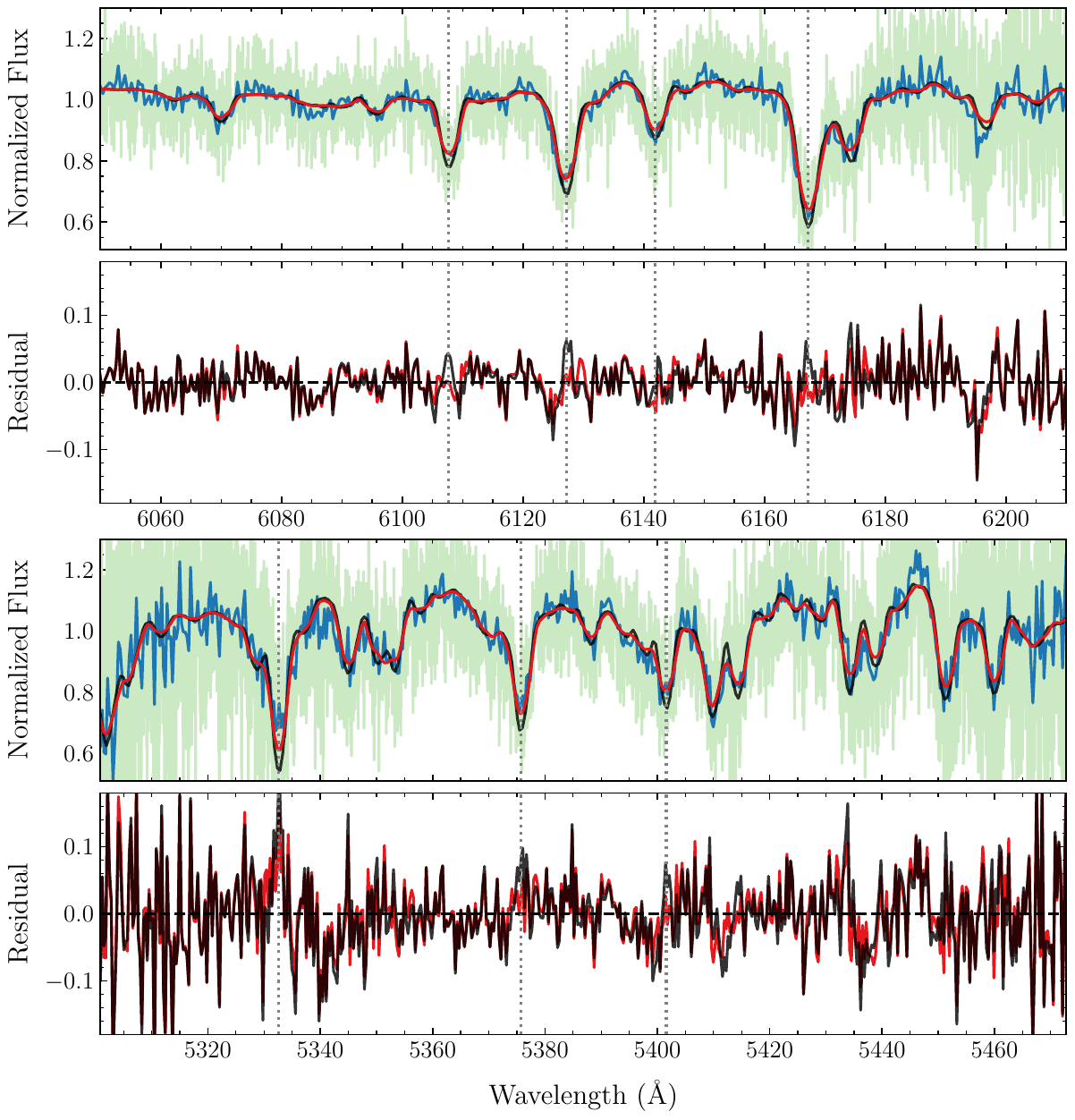}
    \caption{The CFHT spectrum and the best fitting templates. The green 
    spectrum represents the observed data. The blue curves are the rebined 
    spectrum with 10 times of the sampling interval. The red spectrum 
    corresponds to the best-fitting template. The black spectrum is the 
    broadened template using a rotational kernel of $\vsini = 94\,\kms$ 
    \citep{lin2023}. The residuals between the rebined spectrum and the 
    templates are shown in lower panels. Grey vertical dotted lines 
    indicate strong stellar absorption lines. 
    }
    \label{fig:cfht_spec}
\end{figure*}

Figure \ref{fig:cfht_spec} presents two spectral orders of our optimal 
fitting (red curves). For comparison, we fix stellar parameters to their 
best-fitting values and change \vsini\ to 94\,\kms; the resulting broaden 
template is shown as the black curves in Figure \ref{fig:cfht_spec}. The 
chi-square associated with the black curves and the observed data is 
$\chi^2_\mathrm{lin} = 103,877.4$, resulting in a 
reduced-$\chi^2_\mathrm{lin} = 1.302$. It is evident that the red curves 
provide a better match to the observed data than the black curves.

\begin{deluxetable}{lrr}
\tablecaption{The fitting results of CFHT spectrum\label{tab:CFHT_info}}
\tablehead{
\colhead{Parameter} & \colhead{First scheme} & \colhead{Second scheme}
}
\startdata
\teff & 3919 & $4076.8_{-9.7}^{+10.1}$ \\
\logg & $4.81_{-0.03}^{+0.04}$ & $4.98_{-0.02}^{+0.01}$ \\
\feh & $-0.38_{-0.01}^{+0.01}$ & $-0.34_{-0.01}^{+0.01}$ \\
\vsini & $120.7_{-0.9}^{+1.0}$ & $122.5_{-1.0}^{+1.1}$ \\
$\chi^2$ & 103,203.6 & 102,916.4 \\
$\chi^2_\mathrm{lin}$ & 103,994.7 & 103,877.4 \\
\enddata
\tablecomments{The first scheme corresponds to the fitting 
results with a fixed temperature, while the second scheme 
represents the fitting outcomes with temperature as a fitting 
parameter. The final line denotes the chi-squares calculated 
between the observed data and the templates broadened using 
$\vsini_\mathrm{lin} = 94\,\kms$ \citep{lin2023}, with their 
stellar parameters fixed at the values listed in the respective 
columns.}
\end{deluxetable}

The typical SNR of our CFHT spectrum is about 10 (the SNR increases with 
wavelength in the range we use, reaching an SNR of 18 at the red end), 
which is relatively modest. A valid concern is that the SNR may affect 
\vsini\ measurement. To investigate this matter, we resample the CFHT 
spectrum and adjust the sampling interval to be 10 times that of the 
original, thereby increasing the SNR to 32. The resolution of the 
resampled spectrum is reduced to 17,000, which remains sufficiently high 
to preclude any significant broadening effects due to under-sampling. We 
apply the same method to the resampled spectrum,  and the best-fitting 
\vsini\ is consistent with our previous measurements. Hence, our $v\sin i$ 
measurement is robust against data noises. 

\section{Discussion}\label{sec:dis}

The best-fitting effective temperature in our \vsini\ measurement is 
$\teff = 4076.8_{-9.7}^{+10.1}$\,K, slightly larger than the temperature 
($T_\mathrm{eff,lin} = 3919_{-40}^{+31}$\,K) reported by \citet{lin2023}. For 
parameter consistency, we undertake a re-fit of the Spectral Energy 
Distribution (SED) of J1527 utilizing 
\texttt{astroARIADNE}\footnote{\url{https://github.com/jvines/astroARIADNE}} 
\citep{Vines2022}. Here, we establish a prior for the effective temperature 
as $\teff = 4076.8\pm 10$\,K. The radius yielded by this re-fitting is 
$R = 0.637_{-0.004}^{+0.007}\rsun$, slightly less than the radius 
($R_\mathrm{lin} = 0.685_{-0.010}^{+0.016}\,\rsun$) reported in 
\citet{lin2023}.

The mass function for a binary system in a circular orbit is defined as 
follows:
\begin{equation}\label{eq:fm}
    f(M_2) = \frac{M_2^3 \sin^3 i}{(M_1 + M_2)^2} = \frac{K_\mathrm{obs}^3 P_\mathrm{orb}}{2\pi G},
\end{equation}
where $M_1$ and $M_2$ represent the masses of the visible star and the 
compact object, respectively; $i$ denotes the inclination angle; 
$K_\mathrm{obs}$ is the semi-amplitude of the observed RV curve, 
$P_\mathrm{orb}$ is the orbital period, and $G$ stands for the 
gravitational constant. We adopt the mass function of 
$f(M_2) = 0.131\pm 0.002$\,\msun\ reported by \citet{lin2023}. 
For the mass of the visible star, $M_1$, we estimate it using the 
relationship between $K_s$-band absolute magnitude and stellar 
mass provided by \citet{Mann2019}. Based on the apparent magnitude of 
J1527, $\mathrm{Mag}(K_s) = 10.15 \pm 0.02$, and its distance, 
$D = 118.1 \pm 0.1$\,pc, we constrain the mass of the visible star to 
be $M_1 = 0.63 \pm 0.02$\,\msun.

According to the estimated mass and radius of the visible star, 
its filling factor is about 0.92. In this case, the visible star is 
significantly distorted by the gravitational force of the unseen companion. 
Consequently, the projected rotational velocity, $\vsini$, varies with 
the orbital phase, as discussed in 
\citet{Shahbaz1998,Jayasinghe2021,Masuda2021}. Given the notable distortion 
of the visible star, it is inappropriate to use 
$\vrot = 2\pi\,R / P_\mathrm{orb}$ (where $P_\mathrm{orb} = 0.2556698$\,days 
is orbital period) to calculate $V_{\mathrm{rot}}$. Therefore, we use 
\texttt{Phoebe}\footnote{\url{http://phoebe-project.org}} \citep{prvsa2018} 
to compute the rotational velocity at the spectral observation phase 
(phase=0.756) as a given inclination angle. Our steps to determine the 
inclination angle are as follows. First, we start with an arbitrary 
inclination angle ($75$ degrees) and use \texttt{Phoebe} to compute 
$V_{\mathrm{rot}}$. Second, we use the calculated $V_{\mathrm{rot}}$ and 
our $v\sin i$ measurement to obtain a new $i$. Third, we adopt the new 
$i$ and repeat the first and second steps. This iteration stops when the 
inclination angle does not change between the first and third steps. 
We obtain a self-consistent rotational velocity of 
$\vrot = 137.7\pm 2.1\,\kms$. This value corresponds to an inclination 
of $i = 63\pm 2^\circ$, and a mass of the compact object of 
$M_2 = 0.69\pm 0.02\,\msun$. The mass of the compact object is much 
smaller than the result of \citet{lin2023}

Our spectroscopic observation was conducted only at
$\mathrm{phase} = 0.756$, corresponding nearly to the maximum value 
of \vsini. Based on the current fitting parameters, we anticipate the 
\vsini\ measurement to be minimal at phases 0 and 0.5, which is 
$111.2\,\kms$. High-resolution spectroscopic observations at other 
phases could check our measurements and provide a more stringent 
constraint on the orbital inclination.

The light curves of J1527 show significant variations in its profile during 
different observation epochs (see figure 6 in \citealt{lin2023}). Moreover, 
the light curve of each observation can not be described well by a pure 
ellipsoidal variation model. Hence, \citet{lin2023} has to introduce 
significant surface spot activity on the visible star to fit the 
observed light curves. As pointed out by \citet{Luger2021} and 
\citet{Rowan2023}, the inclination angle cannot be well determined if 
the tidally distorted star has star spots. In this case, we speculate 
that the inclination angle cannot be robustly determined by analyzing 
the light curves. 

The $\mathrm{H\alpha}$ emission line of J1527 exhibits features of 
multiple components (see Figure 4 in \citealt{lin2023}). Part of the 
emission line is likely to be caused by the stellar activity of the 
visible star, while the other part may be from the accretion process 
of the companion. Similar observational characteristics have been 
reported in a series of previous works 
\citep[e.g.,][]{Tappert2007,Tappert2011,Parsons2012}. These objects 
are thought to be detached binaries that consist of a WD and a K/M 
companion star almost filling its Roche lobe, thereby qualifying them 
as pre-cataclysmic variables (pre-CVs). The WDs in these pre-CVs 
accrete material from the wind of the K/M companion star, leading to 
additional emission features. \citet{lin2023} argue that if the unseen 
object is a WD, its temperature limit derived from the SED fitting 
corresponds to a lower accretion rate, which is inconsistent with the 
observed $\mathrm{H\alpha}$ emission line, and the system should 
exhibit dwarf nova phenomena. However, the temperature of the WDs in 
these binaries can be notably low; for instance, \citet{Parsons2012} 
reported a binary, SDSS J013851.54-001621.6, with an
$\mathrm{H\alpha}$ emission line from WD and a WD temperature of about
$3600$\,K. J1527 might also be a pre-CV system.

\section{Conclusions}\label{sec:conc}

We have performed a spectroscopic observation of an NS candidate 
reported by \citet{lin2023}, J1527, using the CFHT telescope. The 
high resolution of the CFHT spectrum enables us to make a reliable 
\vsini\ measurement. Through a concurrent fitting of stellar parameters 
and \vsini, we have determined that the projected rotational velocity 
of J1527 is $\vsini = 122.5_{-1.0}^{+1.1}\,\kms$, accompanied by an 
estimated orbital inclination of $63\pm2^\circ$. This result deviates 
significantly from the inclination estimation of about $45^\circ$ 
presented in \citet{lin2023}. Considering that the complex optical 
variation characteristics of J1527 make it problematic to constrain the 
inclination angle, the \vsini\ measurements based on CFHT's high-resolution 
spectrum is more reliable. Based on our inferred orbital inclination, we 
have constrained the mass of the compact object in this system to be 
$M_2 = 0.69\pm0.02\,\msun$, falling within the typical mass range of WD.

\section*{Acknowledgements}

We thank the anonymous referee for constructive suggestions that 
improved the paper. This work was supported by the National Key R\&D 
Program of China under grants 2023YFA1607901 and 2021YFA1600401, 
the National Natural Science Foundation of China under grants
11925301, 12033006, 12103041, 12221003, and 12322303, the Natural 
Science Foundation of Fujian Province of China under grants 2022J06002, 
and the fellowship of China National Postdoctoral Program for Innovation 
Talents under grant BX20230020. Our observation (CTAP2023-A0011; PI: 
Hao-Bin Liu) is kindly supported by China Telescope Access Program (TAP). 
The reported results are based on observations obtained at the 
Canada-France-Hawaii Telescope (CFHT) which is operated by the National 
Research Council of Canada, the Institut National des Sciences de 
l'Univers of the Centre National de la Recherche Scientique of France, 
and the University of Hawaii.

\software{astroARIADNE \citep{Vines2022},
          astropy \citep{2013A&A...558A..33A,2018AJ....156..123A},
          emcee \citep{Foreman2013},
          LBLRTM \citep{Clough1992,Clough2005},
          SciPy \citep{Virtanen2020},
          spectool (\url{https://gitee.com/zzxihep/spectool}),
          stellarSpecModel (\url{https://github.com/zzxihep/stellarSpecModel}),
          Turbospectrum \citep{Plez2012}
          }

\bibliography{main}{}
\bibliographystyle{aasjournal}

\end{document}